# Spreading, Fingering Instability and Shrinking of a Hydrosoluble Surfactant on Water

Saeid Mollaei* and Amir H. Darooneh†

Department of Physics, University of Zanjan, P.O.Box 45196-313, Zanjan, Iran

**ABSTRACT:** We report an experimental investigation of spreading then shrinking of a surfactant droplet at the air-water interface with fingering instability appearing at the edge of the droplet stain. We find out that a droplet of a surfactant on the water shows three regimes of spreading, shrinking and resting at the appropriate parameters values. These regimes can be distinguished by measuring the mean square displacement of the droplet parts, fractal dimension of stain, or radius of fingerless part of the stain (to measure it, we define the angular box-counting dimension), versus time. The shrinking regime is a novel phenomenon.  *Keyword*s: Marangoni spreading, Shrinking, Fractal dimension, Fingering instability.

## 1. Introduction

Over the last years, there has been great interest in Marangoni effect due to the wide class of fundamental and applied problems dealing with. Marangoni flow plays key role in dewetting, spreading and fingering instability[1-7], can be used for self-propulsion[8] and appears in many natural phenomena i.e. liquid thin films stability[9,10] and movement of insects on the water surface[11] and hence has been the subject of extensive investigation.

Marangoni spreading of a surfactant droplet on a thin film and deep liquid layer has been studied widely for decades. Dewetting and fingering instability at the contact line of a spreading surfactant drop and a thin liquid film, also instabilities of Marangoni flow on the thick fluid layer are of the researchers' prime interests[12-15].

Continuous injection of an amphiphile on the surface of a centimeter-thick water layer causes a Marangoni flow by the spreading of the hydrosoluble surfactant, and makes the interface divided in three flow regions: a source (around the injection point), a transparent zone and the outer zone with structures like Rayleigh- Taylor instability patterns[16].

Here we want to report axisymmetric spreading of a surfactant drop on a centimeter-thick water layer, fingering instability rising in the spreading front and shrinking of the rim after spreading runs out.

## 2. Materials and Methods

The experiment is carried out in a cylindrical cuvette of 26 cm diameter. We carefully release the surfactant (butyl glycol-$C_6H_{14}O_2$) drop onto the free surface of water, with zero initial velocity. The drop is seeded with some pigment (we are sure that the pigment do not influence the surface tension) to visualizing the flow. Droplet's volume (which is 10 μl) is controlled by micro-pipette. The water layer thickness is 1 cm and the experiment is under isothermal condition at a temperature of 22°C. Surface tensions of the droplet and water at the experiment temperature are 27.7±0.1 mN/m and 72.7 mN/m respectively. The experiment is recorded by a photo camera at 240 fps looking from above.

When we release the droplet on the water surface, contact of droplet and water-air interface triggers the Marangoni flow and spreading begins. The colorful stain spreads on the surface and part of the droplet deposits to the bottom of the cuvette (density of butyl glycol at the experiment temperature is 0.93 gr/cm$^3$ but, adding some pigment increases the density to 1.16 gr/cm$^3$). Whilst the stain spreads, fingers appear at the rim of the stain and gradually grow. When the stain reaches to the maximum growth of the rim, after a short stop begins to shrink. Shrinking in the rim happens faster than the fingers and it causes fingers get longer. Eventually shrinking runs out and a stain with long fingers emerged at the rim, diffusing slowly remains on the water surface. So we can divide the whole process into three regimes: spreading, shrinking and resting regime.

When we repeat the experiment in a smaller cuvette (of 12 cm diameter), we can see that the wall effects influence the whole process and shrinking does not happen. We also observe the spreading and shrinking of the surfactant droplet on the mm thick layer of glycerol (with a density of 1.23 gr/cm$^3$ at the experiment temperature) in a cuvette with a diameter of 12 cm. Difference between surface tension of the base liquid and droplet may be an important parameter for observing the shrinking phenomenon. The thickness of the water layer, droplet volume and cuvette diameter determine the size of stain and fingers at resting regime. All these are important adjustable parameters in our experiment.

## 3. Results

In this paper, we report the results of the experiment just for the case of spreading and shrinking on the water layer.

To identify the regimes, we plotted the mean square displacement of the droplet parts; $\langle r^2 \rangle$, against time. r is the position of a colored droplet part in the coordinate system fixed at the center of the stain. At each frame of the recorded video the average is computed over the entire colored parts. Reported values in Figure 1 are for a typical experiment. As is seen, in the first regime, $\langle r^2 \rangle$, increases and reaches to a maximum at t ≈ 1.3 s. The second regime begins with decrease in $\langle r^2 \rangle$ and in the third regime, $\langle r^2 \rangle$ remains constant. At the spreading and shrinking regimes we observe power law behavior $\langle r^2 \rangle \sim t^\alpha$ with exponents around α = 0.90 and α = −2.61 respectively.

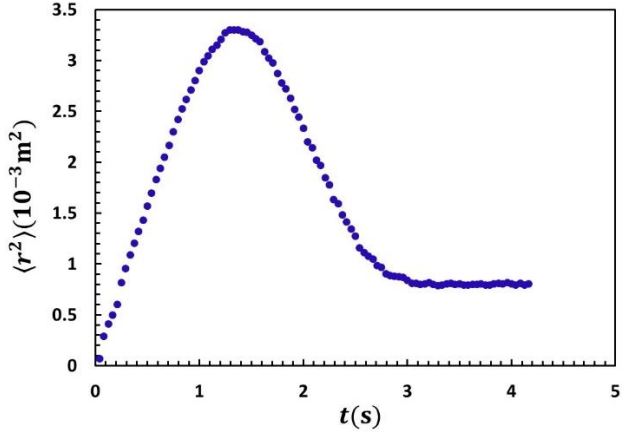

Figure 1. $\langle r^2 \rangle$ versus time. Spreading, shrinking and resting regimes are explicit in the graph.

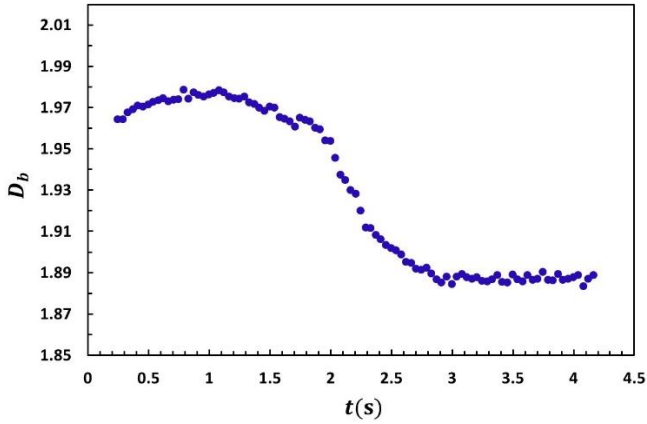

Figure 2. $D_b$ versus time. $D_b$ regime change times compare well with $\langle r^2 \rangle$.

The fractal dimension of stain is close to two, and is constant in the first stage of the experiment. At the end of the first stage, small fingers occur at the stain margin. The fingers become more lengthy and thin in the shrinking phase. As a result the fractal dimension of the stain is reduced. In the final stage the stain is in the resting regime and its shape changes a little then the fractal dimension remains constant again. Figure 2 illustrates the behavior of fractal dimension ($D_b$) in different periods which is in agreement with the results for $\langle r^2 \rangle$.

Radius of the largest circle inscribed in the stain, $r_{rim}$, as a function of time is an alternative quantity to show three stages of spreading, shrinking and resting. For this purpose, we introduce the angular box-counting dimension. The annulus of stain with the radius r from center and a small width $\delta$ can be divided into sectors with the same angle $\theta$ as is illustrated in Figure 3. Each sector that contains a component of the stain is called a filled sector. $N(\theta)$ shows the number of filled sectors in the annulus. The angular box-counting dimension of the annulus is defined as:

$$D_\theta(r) = \lim_{\delta \to 0} \lim_{\theta \to 0} \frac{\log N(\theta)}{\log(\theta)} \quad (1)$$

In practice we choose for $\delta$ a fixed value equal to 10 pixels. $D_\theta$ has a value between 0 and 1, it is equal to one for small r and drops from 1 at $r = r_{rim}$. Then, at the onset of the fingers, $D_\theta$ decreases. Figure 4 exhibits the behavior of $D_\theta$ versus r for three time instances.

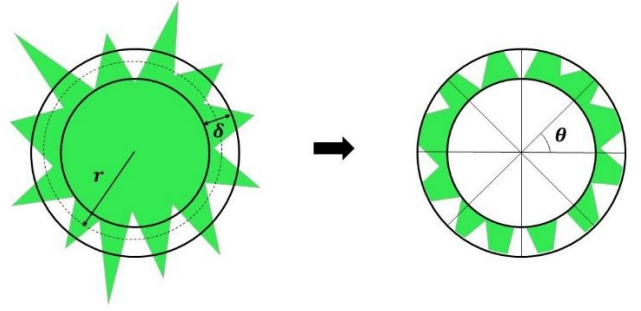

Figure 3. To calculate $D_\theta$, a band of $\delta$ width at the radius r from center of the stain is kept and for different resolution of $\theta$, box-counting is applied.

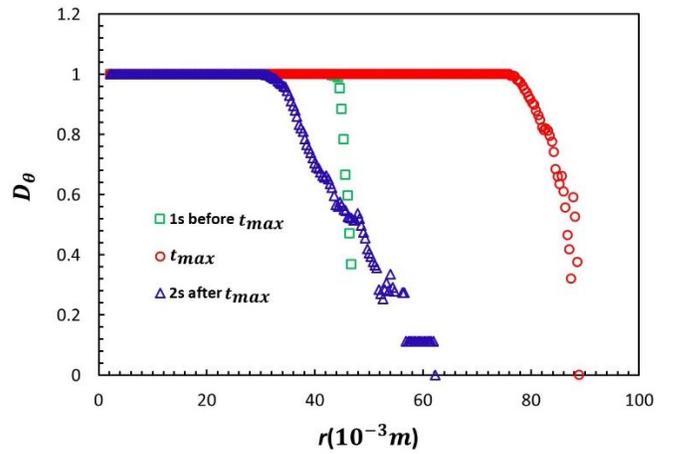

Figure 4. $D_\theta$ versus radial distance from center of the stain for three different times. 2 s after $t_{max}$ the resting regime is reached and fingers are long, hence $D_\theta$ decreases slowly.

In the Figure 5, $r_{rim}$ is computed at each time step and plotted against time for a typical experiment. $r_{rim}$ increases and reaches to a maximum value at the time $t_{max}$, then decreases and finally remains constant, manifesting three regimes claimed above. In the our experiment, $r_{rim}$ is found to have a power law behavior $r_{rim} \sim t^\beta$. It is shown that the spreading exponent ranges from $\beta = 0.25$ to $\beta = 1$[2]. The spreading exponent in the our case is around $\beta = 0.44$, which is close to the exponents reported before by other investigators ($\beta = 0.48$ by Hanyak et al.[4], $\beta = 0.5$ by Hernandez-Sanchez et al.[1] and $\beta = 0.6$ by Starov et al.[3]) for spreading on a thin water layer. For the shrinking regime, the exponent is around $\beta = -1.66$.

We can also see the formation of fingers in Figure 4, the plot of $D_\theta$ versus r for the times 1 s before $t_{max}$, $t_{max}$ and 2 s after $t_{max}$. We see fastest decrease of the $D_\theta$ after corresponding $r_{rim}$, 1 s before $t_{max}$ and slowest 2 s after $t_{max}$. At the beginning of the spreading, fingers do not exist yet or are so small. So $D_\theta$ before $t_{max}$ decreases rapidly. At the $t_{max}$ fingers are long and 2 s after $t_{max}$ we have longest fingers, hence decrease of $D_\theta$ gets slow and slower. Figure 6 shows the shape of stain at 1 s before $t_{max}$, $t_{max}$ and 2 s after $t_{max}$ respectively in binary format and justifies our description.

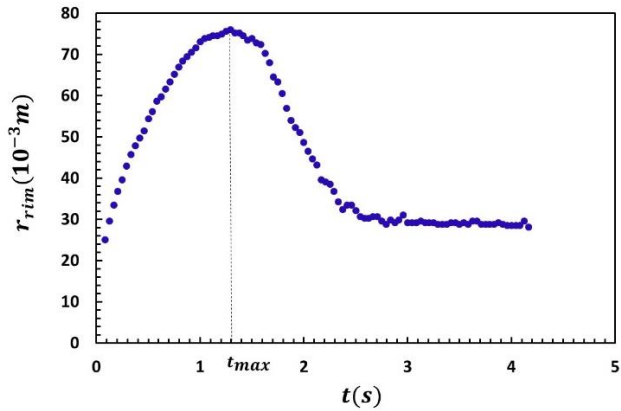

Figure 5. $r_{rim}$ versus time for a typical experiment. Maximum value of $r_{rim}$ occurs at $t_{max}$.

## 4. DISCUSSION AND CONCLUSION

In summary, we have investigated the spreading, shrinking and fingering instability of a surfactant droplet on a thick water layer, experimentally. We have shown that the surfactant droplet during the process experiences three regimes. Also, it has been represented that the fingers, emerging at the edge of the stain start to rise at the first regime and reach to the maximum size at the third regime.

We guess that the spreading droplet on the water surface experiences an elastic-like deformation during the spreading regime, in which crosses the equilibrium point, and at the shrinking regime tries to return to the equilibrium point.

Lately spreading and recession of a liquid droplet during impact on a granular media has been reported[17]. Similar to the shrinking stage in our case, the recession stage of the impacting droplet is an attempt to reach to the equilibrium point of the system. Moreover, one can see resemblance between time evolution diagrams of the diameter of impacting droplet, and radius of the spreading surfactant droplet.

It is observed by Chowdhury et al.[18] that a surfactant-containing water droplet can experience spreading and then recoiling on glass-supported alcohol film. The surface tension of the droplet is larger than the surface tension of the base liquid (alcohol), which disfavors the spreading of the droplet and hence, the phenomenon we have observed is in opposite order of this case.

To find appropriate parameter values for different surfactants to gain a phenomenological and theoretical model of the phenomenon we are performing more experiments on the pure water and glycerol layer.

## AUTHOR INFORMATION

### Corresponding Author

* E-mail: s.mollayi@znu.ac.ir

### Co-author

† E-mail: darooneh@znu.ac.ir

## ACKNOWLEDGMENT

We are grateful to A. Najafi for hosting the experiments at the Complex Fluids Lab of University of Zanjan and to M. Maleki for technical assistance. We also thank R. Shirsavar for helpful discussions.

## REFERENCES

(1) Hernandez-Sanchez, J. F.; Eddy, A.; Snoeijer, J. H. Marangoni spreading due to a localized alcohol supply on a thin water film. *Phys. Fluids* **2015**, 27, 032003.

(2) Matar, O. K.; Craster, R. V. Dynamics of surfactant-assisted spreading. *Soft matter* 2009, 5, 3801-3809.

(3) Starov, V. M.; de Ryck, A.; Velarde, M. G. On the spreading of an insoluble surfactant over a thin viscous liquid layer. *Journal of Colloid and Interface Science* **1997,** 190, 104-113.

(4) Hanyak, M.; Sinz, D. K. N.; Darhuber, A. A. Soluble surfactant spreading on spatially confined thin liquid films. *Soft Matter* **2012,** 8, 7660-7671.

(5) Afsar-Siddiqui, A. B.; Luckham, P. F.; Matar, O. K. Unstable spreading of aqueous anionic surfactant solutions on liquid films. part 1. sparingly soluble surfactant. *Langmuir* **2003,** 19, 696-702.

(6) Nikolov, A.; Wasan, D. Current opinion in superspreading mechanisms. *Adv. Colloid Interface. Sci* **2014**.

(7) Troian, S. M.; Wu, X. L.; Safran, S. A. Fingering instability in thin wetting films. *Phys. Rev. Lett.* **1989**, 62, 1496.

(8) Pimienta, V.; Antoine, C. Self-propulsion on liquid surfaces. *Current Opinion in Colloid & Interface Science* **2014,** 19, 290-299.

(9) Ivanov, I. B.; Kralchevsky, P. A. Stability of emulsions under equilibrium and dynamic conditions. *Colloids and Surfaces A: Physicochem. Eng. Aspects* **1997**, 128, 155-175.

(10) Kelitzing, R. V.; Muller, H. J. Film stability control. *Current Opinion in Colloid & Interface Science* **2002**, 7, 42-49.

(11) Bush, J. W. M; Hu, D. L. Walking on water: biolocomotion at the interface. *Annu. Rev. Fluid Mech.* **2006**, 38, 339-369.

(12) Hamraoui, A.; Cachile, M.; Poulard, C.; Cazabat, A. M. Fingering phenomena during spreading of surfactant solutions. *Colloids and Surfaces A: Physicochem. Eng. Aspects* **2004**, 250, 215-221.

(13) Sinz, D. K. N.; Hanyak, M.; Darhuber, A. A. Immiscible surfactant droplets on thin liquid films: Spreading dynamics, subphase expulsion and oscillatory instabilities. *Journal of Colloid and Interface Science* **2011**, 364, 519-529.

(14) Jensen, O. E. The spreading of insoluble surfactant at the free surface of a deep fluid layer. *J. Fluid Mech.* **1995**, 293, 349-378.

(15) Mizev, A.; Trofimenko, A.; Schwabe, D.; Viviani, A. Instability of Marangoni flow in the presence of an insoluble surfactant. Experiments. *Eur. Phys. J. Special Topics* **2013**, 219, 89-98.

(16) Roche, M.; Li, Z.; Griffiths, I. M.; Le Roux, S.; Cantat, I.; Saint-Jalmes, A.; Stone, H. A. Marangoni flow of soluble amphiphiles. *Phys. Rev. Lett.* **2014**, 112, 208302.

(17) Delon, G.; Terwagne, D.; Dorbolo, S.; Vandewalle, N.; Caps, H. Impact of liquid droplets on granular media. *Phys. Rev. E* **2011**, 84, 046320.

(18) Chowdhury, D.; Sarkar, S. P.; Kalita, D.; Sarma, T. K.; Paul, A.; Chattopadhyay, A. Spreading and recoil of a surfactant-containing water drop on glass-supported alcohol films. *Langmuir* **2004**, 20, 1251-1257.

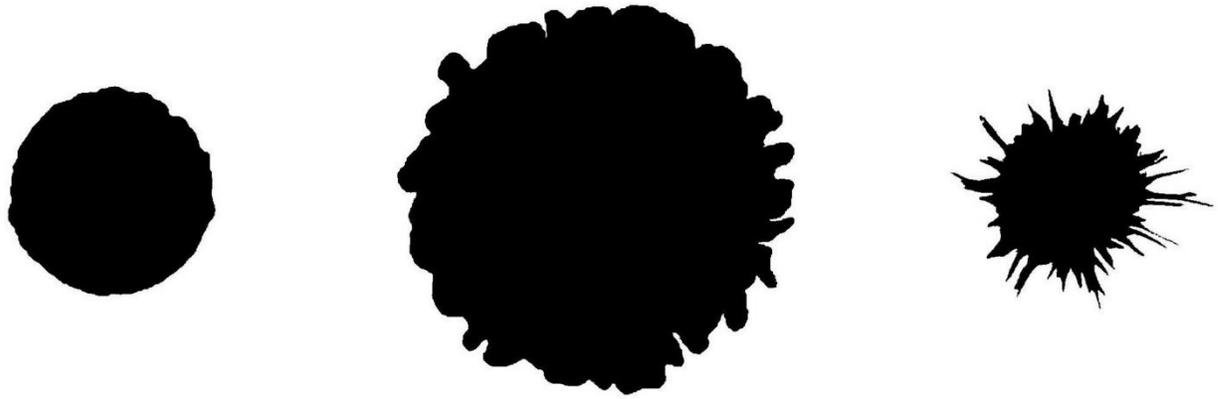

Figure 6. The stain in binary format at three different times: 1 s before $t_{max}$ (left), $t_{max}$ (center) and 2s after $t_{max}$ (right).